\documentclass[%
 reprint,
 amsmath,amssymb,
 aps,
 pra
]{revtex4-1}

\usepackage{graphicx}% Include figure files
\usepackage{dcolumn}% Align table columns on decimal point
\usepackage{bm}% bold math
\usepackage{physics}
\usepackage{verbatim}    % multiline comments
\usepackage[caption=false]{subfig}
\usepackage{xcolor}
\def\sout{\bgroup\markoverwith
{\textcolor{red}{\rule[0.5ex]{2pt}{0.5pt}}}\ULon}
\def\be{\begin{equation}}
\def\ee{\end{equation}}
\def\bes{\begin{equation*}}
\def\ees{\end{equation*}}
\def\bea{\begin{eqnarray}}
\def\eea{\end{eqnarray}}
\def\beas{\begin{eqnarray*}}
\def\eeas{\end{eqnarray*}}
\def\bal#1\eal{\begin{align}#1\end{align}}
\def\bals#1\eals{\begin{align*}#1\end{align*}}

\renewcommand*{\vec}[1]{\boldsymbol{#1}}

\usepackage[normalem]{ulem}

\graphicspath{{figures/}}

\begin{document}

\title{Synthetic spin-orbit coupling mediated by a bosonic environment}

\author{Mikhail Maslov}
\email{mikhail.maslov@ist.ac.at}

\author{Mikhail Lemeshko}%
\email{mikhail.lemeshko@ist.ac.at}
\author{Enderalp Yakaboylu}%
\email{enderalp.yakaboylu@ist.ac.at}
\affiliation{%
 IST Austria (Institute of Science and Technology Austria), Am Campus 1, 3400 Klosterneuburg, Austria
}%

\date{\today}% It is always \today, today,

\begin{abstract}

We study a mobile quantum impurity, possessing internal rotational degrees of freedom, confined to a ring in the presence of a many-particle bosonic bath. By considering the recently introduced rotating polaron problem, we define the Hamiltonian and examine the energy spectrum. The weak-coupling regime is studied by means of a variational ansatz in the truncated Fock space. The corresponding spectrum indicates that there emerges a coupling between the internal and orbital angular momenta of the impurity as a consequence of the phonon exchange. We interpret the coupling as a phonon-mediated spin-orbit coupling and quantify it by using a correlation function between the internal and orbital angular momentum operators. The strong-coupling regime is investigated within the Pekar approach and it is shown that the correlation function of the ground state shows a kink at a critical coupling, that is explained by a sharp transition from the non-interacting state to the states that exhibit strong interaction with the surroundings. The results might find applications in such fields as spintronics or topological insulators, where spin-orbit coupling is of crucial importance.

\end{abstract}

\maketitle

%\tableofcontents

\section{Introduction}

The basic theory of the spin-orbit coupling (SOC) is discussed in most conventional quantum mechanics textbooks \cite{Landau-Quantum,Griffiths-Quantum,Sakurai-Advanced}. SOC is usually treated as an interaction between the magnetic dipole moment of the electron, associated with its intrinsic spin and the magnetic field  in the rest frame of the electron induced by the positively charged nucleus~\cite{Thomas-SOcoupling}. Moreover, an effective spin-orbit coupling can be engineered by using an intense external laser field~\cite{SO-laser-1,SO-laser-2,SO-laser-3}. Apart from leading to splittings in atomic spectra \cite{Russel-Saunders} and being one of the main ingredients of the nuclear shell model \cite{Iwanenko,Mayer}, SOC was shown to be of great importance in a variety of condensed-matter phenomena. To name a few, it underpins the Dresselhaus \cite{Dresselhaus} and Rashba \cite{Rashba,Bychkov-Rashba} splittings that were used in spintronics to create a spin-injected field transistor \cite{SFET,Rashba-spintronics,Spintronics-review} and antiferromagnetic memory \cite{AFM-SO,AFM-memory,AFM-review}. It also stands behind the spin Hall effect -- spin imbalance generated by charge circulation in paramagnets \cite{Dyakonov-Perel,Hirsch,Murakami-SH,Sinova-SH} as well as its quantum counterpart \cite{Kane-Mele-QSH} that leads to the distinction of topological phases \cite{Kane-Mele-topology}, giving rise to the discussions on topological insulating electronic phases \cite{TI-1,TI-2,TI-3} and topological quantum computation \cite{Kitaev,Freedman,Sarma}, e.g. using Majorana bound states \cite{Kane-Majorana}.  Achieved by coupling internal atomic degrees of freedom with laser fields \cite{gauge-cold-atoms}, the spin-orbit interaction has also been realised in ultracold gases \cite{SO-ultracold,SO-ultracold-nature} with topologically nontrivial states emerging due to the interaction with synthetic gauge fields \cite{ultracold-topology}.

Here we consider another kind of spin-orbit coupling, i.e., that induced by the excitations of a many-particle environment. In particular, we focus on a bosonic many-particle  environment, acting as a mediator of the interaction between two angular momenta corresponding to the internal and orbital motion of an impurity in the bath. Such motion may be achieved, e.g., by applying a static magnetic field to a system consisting of charged molecule and neutral environment. For the impurity we consider a quantum rotor. A rotating impurity surrounded by a bosonic bath was previously shown to be effectively represented by a quasiparticle named ``angulon''~\cite{PRL-angulon,PRX-angulon,Angulon-chapter}, which can be described by using the language of Feynman diagrams for the excitations in the phonon continuum~\cite{PRB-giacomo,PRL-giacomo}. It has been also shown that the angulon quasiparticle leads to novel phenomena such as realization of magnetic monopoles in terms of molecular impurities~\citep{PRL-gauge} or strong ``anomalous'' electromagnetic screening~\citep{PRL-screening}. Moreover, the angulon theory can explain the anomalous broadening of the spectroscopic lines observed in superfluid helium nanodroplets~\citep{PRM-igor}.
 
The paper is organized as follows. In Section \ref{sec:hamiltonian}, we present the Hamiltonian of the system obtained from the recently introduced ``rotating polaron'' quasiparticle \citep{PRB-polaron} by taking into account the geometrical aspects of our system. We subsequently introduce a coordinate transformation that brings the Hamiltonian into the reference frame co-rotating with the impurity's center-of-mass, eliminating the angle of circulation from the Hamiltonian. In Section \ref{sec:weak-coupling}, we investigate the Hamiltonian by using one-phonon variational ansatz in the regime of the weak coupling between the impurity degrees of freedom and the many-body environment. We present the energy spectrum of the system and illustrate the transition between different angular momentum states. The transition is explained within the framework of the spin-orbit coupling by calculating the correlators between the angular momentum operators. We show that in the vicinity of the angulon instability~\cite{PRL-angulon} the coupling amplifies due to the increased probability of the phonon exchange. In Section \ref{sec:strong-coupling} we follow the Pekar approach \cite{Pekar} to examine the effect in the strong-coupling limit and thereby establish a connection between the bath-mediated angular momentum exchange and the effective spin-orbit interaction of the molecule in the medium. We conclude the paper in Sec.~\ref{sec_conc} with a discussion of our results.

\section{The Hamiltonian}
\label{sec:hamiltonian}
\begin{figure}
    \centering
    \includegraphics[width=8.6cm]{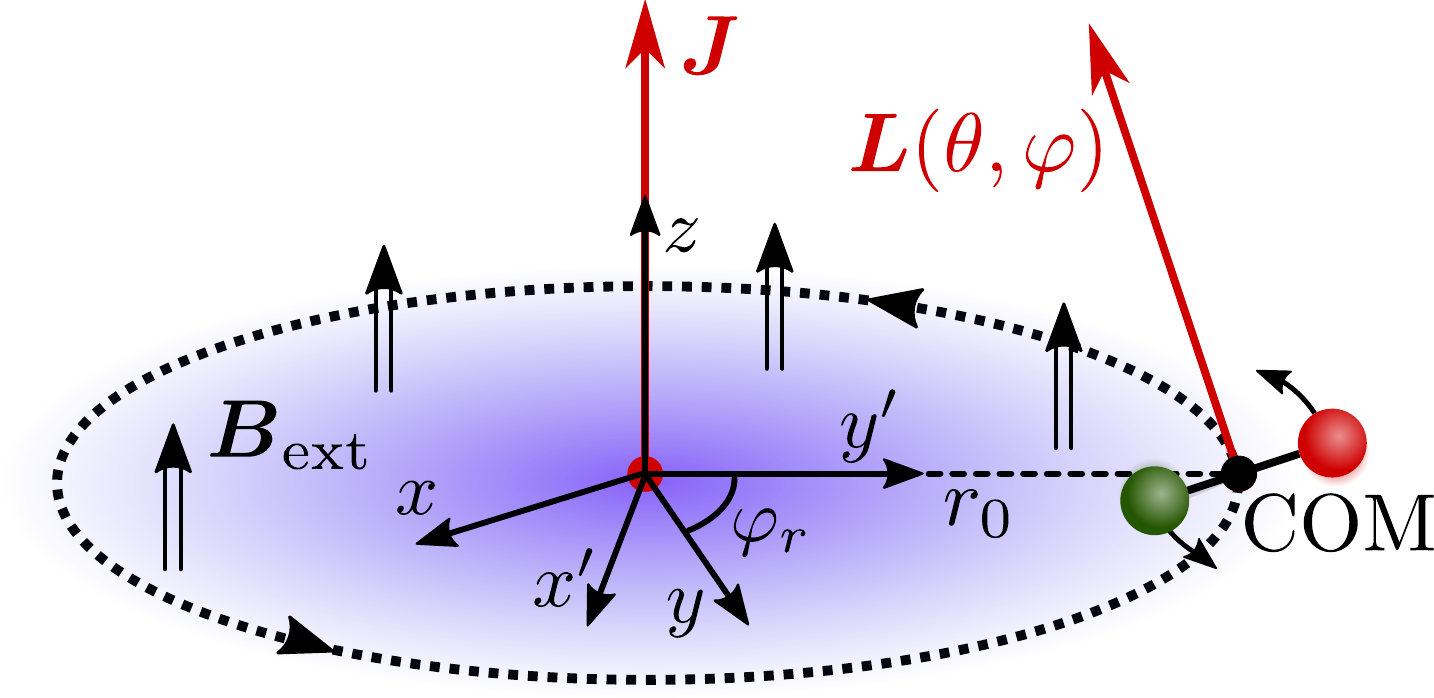}
    \caption{Schematic diagram of a rotating impurity immersed into the bosonic bath. The center-of-mass (COM) transverse motion is confined to a ring of radius $r_{0}$ in the $xy$-plane (dashed line). Here $(x,y,z)$ and $(x^{\prime},y^{\prime},z^{\prime})$ are the laboratory and circulation frames, respectively.}
    \label{fig:system}
\end{figure}

The Hamiltonian that describes a moving impurity posessing internal rotational degrees of freedom immersed in a bosonic environment, derived from the first principles within the Bogoliubov approximation \cite{Bogoliubov}, was studied in Ref. \cite{PRB-polaron}. Here we consider the case, where the transverse motion of the impurity's center-of-mass (COM) is confined to a ring of radius $r_{0}$ in the $xy$-plane. The geometry of the problem is sketched in Fig.~\ref{fig:system}. Such a configuration can be obtained for example by considering a charged molecule in a neutral bath subjected to an external magnetic field $\vec{B} = B_{\text{ext}} \hat{z}$. Then, as it is described in more detail in Ref.~\cite{PRB-polaron}, the corresponding Hamiltonian for this configuration is given by
\bal
    \label{eq:hamiltonian}
    \hat{\mathcal{H}}& = B_{\text{ext}} (\hat{J}_{z}+\hat{L}_{z})+B_{0}\hat{J}_{z}^{2}+B\hat{\bm{L}}^{2} \\
\nonumber &+\sum\limits_{k \lambda \mu}\omega(k)\hat{b}_{k \lambda \mu}^{\dagger}\hat{b}_{k \lambda\mu}+\hat{\mathcal{H}}_{\text{int}} \, ,
\eal
%\begin{multline}
%    \hat{\mathcal{H}}=B_{\text{ext}}(\hat{J}_{z}+\hat{L}_{z})+B_{0}\hat{J}_{z}^{2}+B\hat{\bm{L}}^{2}\\
%    +\sum\limits_{\bm{k}}\omega(k)\hat{b}_{\bm{k}}^{\dagger}\hat{b}_{\bm{k}}+\hat{\mathcal{H}}_{\text{int}},
% %   \label{eq:hamiltonian}
%\end{multline}
where atomic units are used and $B_{0}\equiv(2\mathcal{M}r_{0}^{2})^{-1}$, $\mathcal{M}$ is the mass of the molecular impurity, $B$ the molecular rotational constant, $\hat{\bm{J}}$ the angular momentum of circular motion, and $\hat{\bm{L}}$ the angular momentum of the impurity's internal rotation. The first term in Eq.~\eqref{eq:hamiltonian} describes the interaction between the angular momentum of the molecular impurity with the external magnetic field $B_{\text{ext}}$, the second stands for the kinetic energy of the COM motion, whereas the third is the kinetic energy of the internal rotation. 
The forth term, in turn, corresponds to the kinetic energy of the bath with $\sum_{k} \equiv \int d k$. Here $\hat b^\dagger_{k \lambda \mu}$ and $\hat b_{k \lambda \mu}$ are the bosonic creation and annihilation operators written in the spherical basis. They obey the commutation relation $[\hat b_{k \lambda \mu}, \hat b^\dagger_{k' \lambda' \mu'}] = \delta(k-k')\delta_{\lambda \lambda'} \delta_{\mu \mu'}$. Finally, the last term describes the impurity-phonon interaction,
\begin{equation}
    \hat{\mathcal{H}}_{\text{int}}=\sum\limits_{\substack{k\lambda\mu \\ l\delta\alpha\gamma}}\Big[\mathcal{U}_{l\alpha\lambda}^{\delta\gamma\mu}(k)j_{l}(kr_{0})\hat{Y}_{l,\delta}^{\ast}(\hat{\Omega}_{r})\hat{Y}_{\alpha,\gamma}^{\ast}(\hat{\Omega})\hat{b}^{\dagger}_{k\lambda\mu}+H.c.\Big] \, ,
    \label{eq:interaction}
\end{equation}
with $\hat{r}\equiv(r_{0},\hat{\Omega}_{r})\equiv(r_{0,}\frac{\pi}{2},\hat{\varphi}_{r})$ being the circulation operators of the COM coordinate, $\hat{\Omega}\equiv(\hat{\theta},\hat{\varphi})$ angular operators of internal rotation, $j_{l}(kr_{0})$ the spherical Bessel function of the first kind, and $\hat{Y}_{l,m}(\hat{\Omega})$ the spherical harmonics. The coupling amplitude is given by
\begin{equation}
    \mathcal{U}_{l\alpha\lambda}^{\delta\gamma\mu}(k)=U_{\alpha}(k)\text{\footnotesize $\sqrt{\frac{4\pi(2\alpha+1)(2l+1)}{2\lambda+1}}$} i^{\lambda-\alpha-l}C_{\alpha 0,l0}^{\lambda 0}C_{\alpha\gamma,l\delta}^{\lambda\mu} \,,
\end{equation}
where $C_{\alpha\gamma,l\delta}^{\lambda\mu}$ are the Clebsch-Gordan coefficients \cite{Varshalovich}.

Following Ref.~\citep{PRL-angulon}, we derive the interaction term~\eqref{eq:interaction} by considering a local density-density interaction between the impurity and bosonic surroundings. Therefore, the conservation of the total angular momentum in the system is embedded in Eq.~\eqref{eq:interaction} by the presence of the aforementioned Clebsch-Gordan coefficients. Moreover, the interaction term of the Hamiltonian~\eqref{eq:hamiltonian} already hints the mechanism of the interaction between the external and internal angular momenta of the impurity. In particular, it couples these angular momenta together through the angular momentum of a bosonic bath, which signals that the SOC is mediated by the phonon excitations in the bath.

It is straightforward to show that the derived Hamiltonian is rotationally invariant. In other words, the total angular momentum of the system, $\hat{K}=\hat{J}_{z}+\hat{L}_{z}+\hat{\Lambda}_{z}$, is a constant of motion, i.e., it commutes with the Hamiltonian. The latter indicates that the energy eigenstate can be constructed from the eigenstate of the total angular momentum, which will be discussed below in a more detail way. Furthermore, it allows us to consider the problem in the reference frame co-rotating with the molecule's center-of-mass ($x^{\prime},y^{\prime},z^{\prime}$). Namely, if we introduce the following canonical transformation
\begin{equation}
    \hat{S}(\hat{\varphi}_{r})=\exp\Big[-i\hat{\varphi}_{r}\big(\hat{\Lambda}_{z}+\hat{L}_{z}\big)\Big] \,,
\end{equation}
with $\hat{\bm{\Lambda}}=\sum\limits_{k\lambda\mu\nu}\bm{\sigma}^{\lambda}_{\mu\nu}\hat{b}^{\dagger}_{k\lambda\mu}\hat{b}_{k\lambda\nu}$ being the collective angular momentum of the bosonic bath, the transformed Hamiltonian becomes independent from the external rotational angle, $\hat{\varphi}_{r}$,
\begin{multline}
    \hat{\mathcal{H}}^{\prime}\equiv\hat{S}^{-1}\hat{\mathcal{H}}\hat{S}
    =B_{\text{ext}}\big(\hat{J}_{z}-\hat{\Lambda}_{z}\big)+B_{0}\big(\hat{J}_{z}-\hat{L}_{z}
    -\hat{\Lambda}_{z}\big)^{2}\\+B\hat{\bm{L}}^{2}
    +\sum\limits_{k \lambda \mu}\omega(k)\hat{b}_{k \lambda \mu}^{\dagger}\hat{b}_{k \lambda}+\hat{\mathcal{H}}_{\text{int}}^{\prime} \,.
    \label{eq:transformed-hamiltonian}
\end{multline}
Here the molecule-bath coupling term is given by
\begin{equation}
    \hat{\mathcal{H}}_{\text{int}}^{\prime}=\sum\limits_{k\lambda\mu}\sum\limits_{\alpha\gamma}\big[V_{\alpha\lambda}^{\gamma\mu}(k)Y_{\alpha,\gamma}^{\ast}(\hat{\Omega})\hat{b}_{k\lambda\mu}^{\dagger}+H.c.\big] \,,
\end{equation}
with the interaction amplitude
\begin{equation}
    V_{\alpha\lambda}^{\gamma\mu}(k)=\sum\limits_{l\delta}\mathcal{U}^{\delta\gamma\mu}_{l\alpha\lambda}(k)j_{l}(kr_{0})Y_{l,-\delta}\Big(\frac{\pi}{2},0\Big) \,.
\end{equation}

The total angular momentum of the system in the co-rotating frame, on the other hand, is equal to the angular momentum of the circular motion
\begin{equation}
    \hat{S}^{-1}(\hat{J}_{z}+\hat{L}_{z}+\hat{\Lambda}_{z})\hat{S}=(\hat{J}_{z}-\hat{L}_{z}
    -\hat{\Lambda}_{z})+\hat{L}_{z}+\hat{\Lambda}_{z}=\hat{J}_{z} \, .
    \label{eq:total-angular-momentum}
\end{equation}
Therefore, in the co-rotating frame the eigenvalues of the angular momentum $\hat{J}_{z}$ label the total angular momentum numbers.

\section{Weak-coupling regime}
\label{sec:weak-coupling}

Together with the conservation of the total angular momentum, Eq. (\ref{eq:total-angular-momentum}) results in a constraint on the trial wavefunction: in the transformed frame it should be an eigenstate of the $\hat{J}_{z}$ operator. Therefore, one can write down such a wavefunction with the explicit dependence on the external rotation angle, $\hat{\varphi}_{r}$,
\begin{equation}
    \bra{\varphi_{r}}\hat{S}(\hat{\varphi}_{r})\ket{\psi}=\frac{1}{\sqrt{2\pi}}\exp[iM\varphi_{r}]\hat{S}(\varphi_{r})\ket{\psi_{M}} \, ,
    \label{eq:trial-wavefunction}
\end{equation}
where $M$ is the eigenvalue of total angular momentum operator.

% \oldtext{Let us assume that the interaction between the molecular impurity and the many-particle environment is weak compared to the molecular kinetic energy, parametrized by $B$.}

The wavefunction (\ref{eq:trial-wavefunction}) is a tensor product of circular motion eigenstate and the eigenstate of a steady impurity coupled to a bath. Let us assume that the characteristic energy of the interaction between the molecular impurity and the many-particle environment is small compared to either the molecular kinetic energy, parametrized by $B$ and $B_{0}$, or the kinetic energy of the bath. In this case, we can consider the Fock space alongside with truncating the number of phonons. In order to solve the eigenvalue equation of the Hamiltonian (\ref{eq:transformed-hamiltonian}) in this regime, we introduce the following single-phonon variational ansatz (similar to Ref. \cite{PRL-angulon})
\begin{equation}
    \ket{\psi_{M}}=g\ket{LM_{L}}\ket{0}+\sum\limits_{k\lambda\mu jm}\alpha_{k\lambda\mu jm}\ket{jm}\hat{b}^{\dagger}_{k\lambda\mu }\ket{0} \,,
    \label{eq:ansatz}
\end{equation}
where $L$ and $M_{L}$ are eigenvalues of $\hat{\bm{L}}^{2}$ and $\hat{L}_{z}$ operators, respectively, and $g^{\ast}$ and $\alpha_{k\lambda\mu jm}^{\ast}$ are variational parameters. The minimization of the energy functional $\mathcal{F}=\bra{\psi}\hat{H}-E\ket{\psi}$ with respect to parameters $g^{\ast}$ and $\alpha_{k\lambda\mu jm}^{\ast}$ results in the self-consistent Dyson equation
\begin{equation}
    BL(L+1)+B_{\text{ext}}(M_{J}+M_{L})+B_{0}M_{J}^{2}-E-\Sigma^{M_{J}}_{LM_{L}}(E)=0 \,,
    \label{eq:dyson}
\end{equation}
with $M_{J}\equiv M-M_{L}$ being the eigenvalue of the angular momentum of circular motion. The self-energy is given by
\begin{multline}
    \Sigma_{M_{J},LM_{L}}(E)=\sum\limits_{k\lambda\mu jm}\big|\xi_{\lambda\mu,jm}^{LM_{L}}(k)\big|^{2}\Big(B_{\text{ext}}(M_{J}+M_{L}-\mu)\\+B_{0}(M_{J}+M_{L}-m-\mu)^{2}+Bj(j+1)+\omega(k)-E\Big)^{-1} \, ,
\end{multline}
where
\begin{equation}
    \xi_{\lambda\mu,jm}^{LM_{L}}(k)=\sum\limits_{\alpha\gamma}V_{\alpha\lambda}^{\gamma\mu}(k)\sqrt{\frac{(2\alpha+1)(2j+1)}{4\pi(2L+1)}}C_{\alpha 0,j0}^{L0}C_{\alpha\gamma,jm}^{LM_{L}} \,.
\end{equation}

The energies for each set of the free parameters $\{L,M_{L},M_{J}\}$ that characterizes the state may be found self-consistently via the poles of the Green's function
\begin{equation}
    G_{M_{J},LM_{L}}(E)=\frac{1}{BL(L+1)-\Sigma_{M_{J},LM_{L}}(E)-E} \,,
\end{equation}
and the entire excitation spectrum of the system may be examined via the spectral function,
\begin{equation}
	\mathcal{A}_{M_{J},LM_{L}}=\Im[G_{M_{J},LM_{L}}(E+i0^{+})] \, .
	\label{eq_spec_func}
\end{equation}

In order to provide a quantitative description of the spectral function, we adapt the parameters from Ref. \cite{PRL-angulon}, such as the Bogoliubov dispersion relation
\begin{equation}
    \omega(k)=\sqrt{\varepsilon(k)(\varepsilon(k)+2g_{\text{bb}}n)} \,  , 
\end{equation}
where the boson-boson interaction is considered constant $g_{\text{bb}}=4\pi a_{\text{bb}}/m_{b}$ and the boson-boson scattering length $a_{\text{bb}}=3.3/\sqrt{m_{b}B}$, as well as the impurity-boson interaction potential
\begin{equation}
    U_{\lambda}(k)=\sqrt{\frac{8nk^{2}\varepsilon(k)}{\omega(k)(2\lambda+1)}}\int drr^{2}f_{\lambda}(r)j_{\lambda}(kr) \,,
    \label{eq:interaction-potential}
\end{equation}
with the Gaussian form factor $f_{\lambda}(r)=u_{\lambda}(2\pi)^{3/2}e^{-r^{2}/(2r_{\lambda}^{2})}$ and the parameters $\lambda=\{0,1\}$, $u_{0}=1.75 u_{1}=218B$, and $r_{0}=r_{1}=1.5/\sqrt{m_{b}B}$. These parameters are chosen to approximate the range and strength of a typical atom-molecule potential.

\begin{figure}
   \centering
	\includegraphics[width=8.6cm]{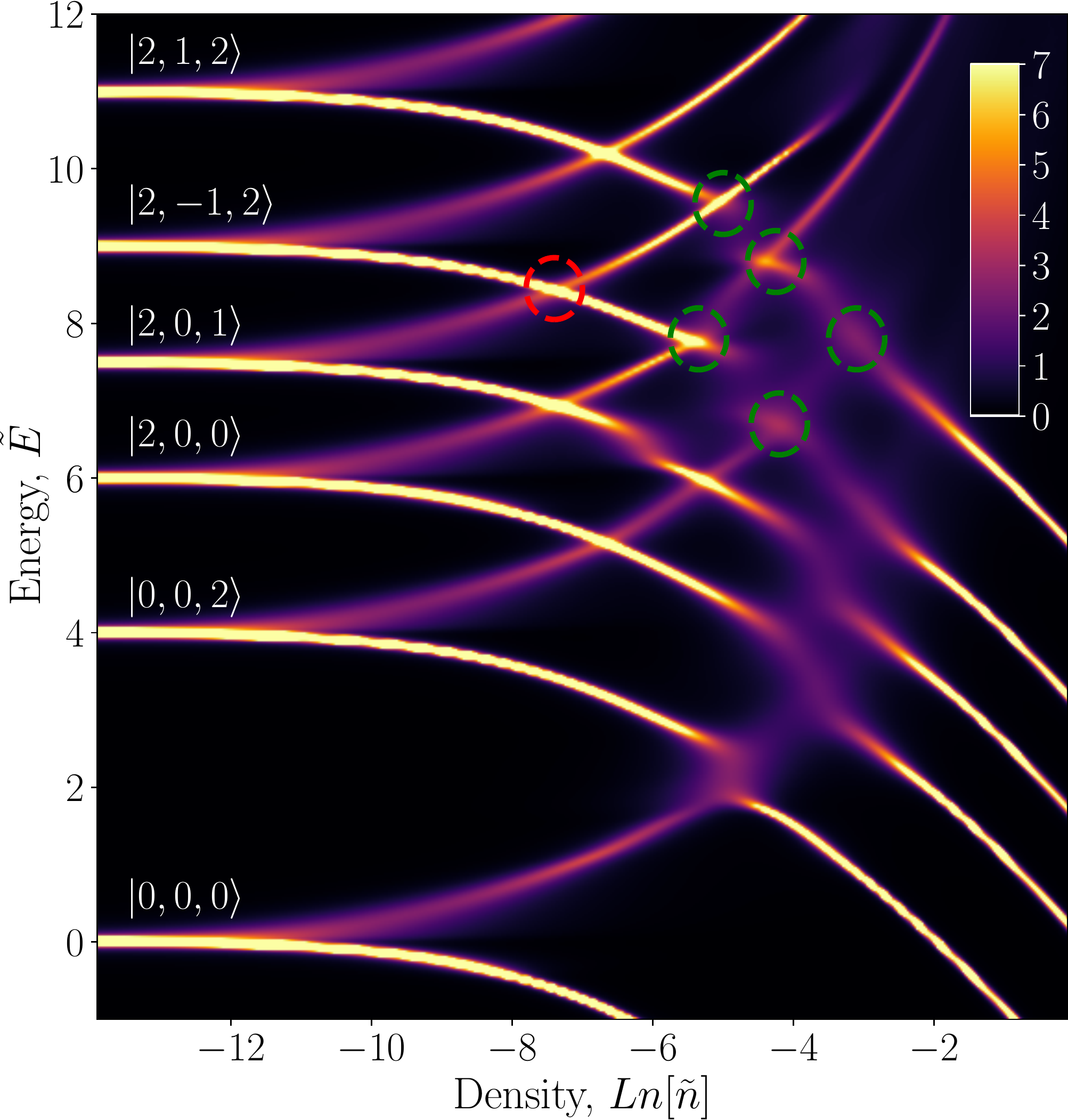}
	\vspace{-0.5\baselineskip}
    \caption{The spectral function of the system, $\mathcal{A}_{M_{J},LM_{L}}$, as a function of  the normalized energy and bath density for different configurations of the free parameters $\{L,M_{L},M_{J}\}$. The state $\ket{2,1,2}$ exhibits an additional transition from the $\ket{2,0,1}$ state (green circles), whereas the same state with the inverted orientation of internal rotation $\ket{2,-1,2}$ does not reveal such a transition (red circle).}
    \label{fig:spectrum}
\end{figure}

In Fig.~\ref{fig:spectrum} the spectral function~\eqref{eq_spec_func} as a function of the normalized energy, $\tilde{E}=E/B$ and normalized bath density, $\tilde{n}=n(m_{b}B)^{3/2}$, is presented. With the increasing density, as the rotor and its surroundings begin to effectively interact, system's energy levels split into two branches. The bright and sharp lines correspond to stable quasiparticle states that exhibit a negative shift in the energy, and they are dressed by bosonic excitations in high-density regime. The darker blurred regions, on the other hand, depict the metastable excited states of the phonon continuum that become completely decoherent at large values of $\tilde{n}$ due to the spontaneous emission to phonon vacuum. When the quasiparticle states intersect with the phonon excitations, the angular momentum is being transfered from the internal degrees of freedom to the bath \cite{PRL-angulon}. The transition occurs only if the intersection reveals the disintegration of the stable quasiparticle state (green circles), later referred to as instability, and does not occur in other cases (red circle).

\begin{figure}
   \centering
	\includegraphics[width=8.6cm]{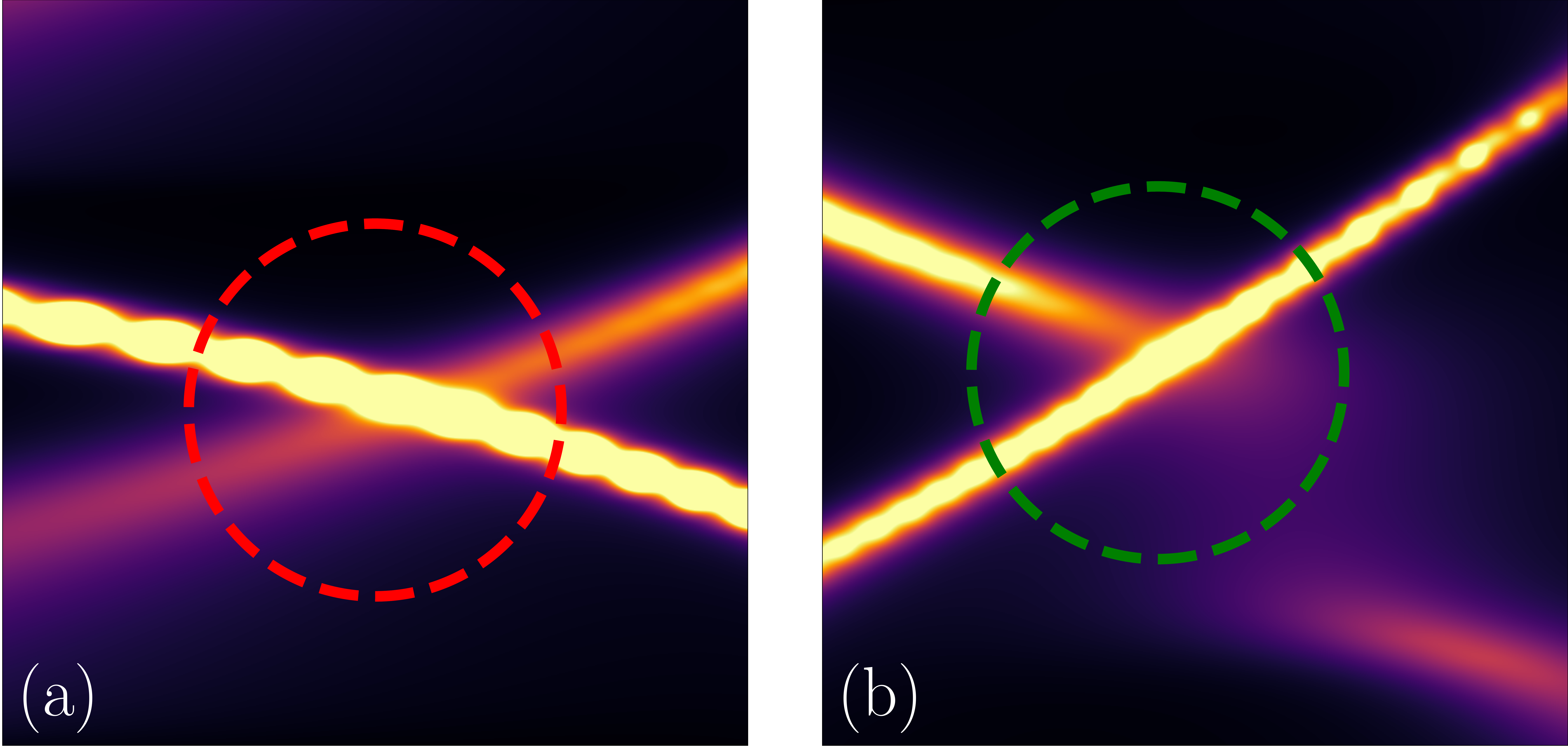}
	\vspace{-0.5\baselineskip}
    \caption{The enlarged regions of Fig.~\ref{fig:spectrum} in the vicinity of the crossings between the metastable state of the phonon continuum associated with the impurity state $\ket{2,0,1}$ and stable quasiparticle branches of states $\ket{2,-1,2}$ (a) and $\ket{2,1,2}$ (b). From the behavior of spectral function, one can deduce the information about the suppressed (a) and allowed (b) transitions.}
    \label{fig:spectrum-allowed}
\end{figure}

We picture the spectral function for those specific states that permit the coupling between external and internal momenta through the excitations in many-body surroundings. For these states the behavior of the spectral function in the vicinity of instability regions is governed by the values of quantum numbers $M_{J}$ and $M_{L}$ with respect to each other. We will further use the notation $\ket{L,M_{L},M_{J}}$ to define the state. In Fig.~\ref{fig:spectrum},  the state $\ket{2,-1,2}$ has only two allowed transitions, with lower levels: states $\ket{2,0,0}$ and $\ket{0,0,2}$. The state with the inverted orientation of internal rotation, $\ket{2,1,2}$, on the other hand, exhibits an additional transition from $\ket{2,0,1}$ energy level. In Fig.~\ref{fig:spectrum-allowed}, we present the enlarged images of the transition regions. In the regime, where the angular momentum transfer does not occur (see Fig.~\ref{fig:spectrum-allowed}(a)) the quasiparticle state $\ket{2,-1,2}$ maintains its stability by crossing the phonon branch of state $\ket{2,0,1}$. However, the quasiparticle state $\ket{2,1,2}$ disintegrates in the vicinity of the crossing (see Fig.~\ref{fig:spectrum-allowed}(b)), which reveals the angular momenta redistribution according to Ref.~\citep{PRL-angulon}. By observing such a behavior of the spectral function, one may suggest that the transition probability should depend on the relative projections of the participating angular momenta. In general, that would imply that the energy of each state is dependent on the coupling between external and internal angular momenta.

The quantitative evaluation of such coupling, however, is not that straightforward. Provided that the interaction between two angular momenta is mediated by the phonon exchange, we choose the correlation of $\hat{J}_{z}$ and $\hat{L}_{z}$ operators as a measure of the coupling. The energy solution to Eq.~\eqref{eq:dyson} together with the normalization condition on the state (\ref{eq:ansatz}) allows us to obtain values of the variational coefficients $g$ and $\alpha_{jmk\lambda\mu}$, and thereby numerically defines the trial state (\ref{eq:ansatz}). With the wavefunction of the system being numerically defined for each set of $\{L,M_{L},M_{J}\}$, one can further calculate the expectation value via the following relation
\begin{equation}
    \mathcal{C}\equiv\bra{\psi}\hat{J}_{z}\hat{L}_{z}\ket{\psi}-\bra{\psi}\hat{J}_{z}\ket{\psi}\bra{\psi}\hat{L}_{z}\ket{\psi} \,.
    \label{eq:correlation}
\end{equation}

\begin{figure}
    \centering
    \includegraphics[width=8.6cm]{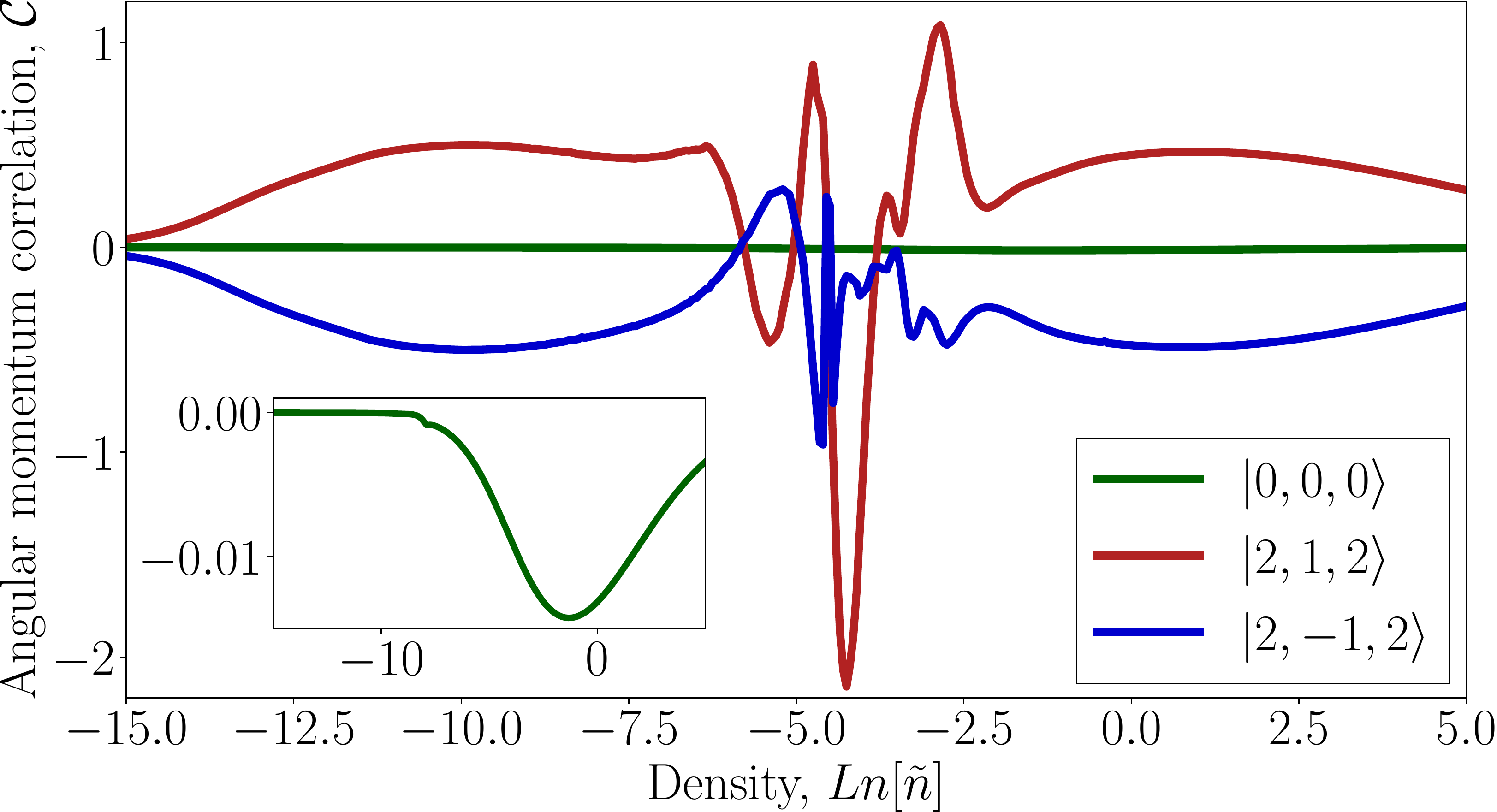}
    \vspace{-0.5\baselineskip}
    \caption{The dependence of angular momentum correlation on the density of the bosonic bath. In the ground state, $\ket{0,0,0}$, the $\hat{J}_{z}$ and $\hat{L}_{z}$ operators exhibit infinitesimally small correlation (inset), whereas in the excited states the two are coupled to each other in the instability regime with the maximum amplitude dependent on the relative projections of momenta.}
    \label{fig:correlation}
\end{figure}

The dependence of the coupling strength on the normalized bath density, calculated within Eq.~(\ref{eq:correlation}) is presented in Fig.~\ref{fig:correlation}. We consider three sets of angular momentum projections: $\ket{0,0,0}$ - corresponding to the ground state of the system, where the correlation is infinitesimally low, however not zero, as shown on the inset, and $\ket{2,-1,2}$ and $\ket{2,1,2}$, for which the results agree with the behavior of the spectral function in Fig.~\ref{fig:spectrum}. Nonzero values for the ground state (inset) are explained by the fact of spontaneous creation of a phonon in the system with zero initial rotations.  Due to the conservation of total angular momentum such a phonon should induce both rotations, thus actuating the interaction between them. The correlation in general is stronger for the positive values of $M_{L}$, given positive values of $M_{J}$, which allows us to draw a parallel to a conventional term in the relation for the energy representing the spin-orbit coupling: $\propto\hat{\bm{J}}\cdot\hat{\bm{L}}$. We therefore conclude that the selective transition behavior illustrated in Fig.~\ref{fig:spectrum} and Fig.~\ref{fig:spectrum-allowed} is a consequence of the interaction between the internal (``spin'') and external (``orbit'') rotational degrees of freedom of the impurity.

The results illustrated on Fig.~\ref{fig:spectrum} and Fig.~\ref{fig:correlation} are also valid in the limit $B_{\text{ext}}\rightarrow 0$. This means that the observed effect is only a consequence of the interaction between two angular momenta and is enhanced, but not created, by the external magnetic field.

Ultimately, the transitions depicted in Fig.~\ref{fig:spectrum} may be described by a set of selection rules. The prohibited ones are those preserving the quantum numbers $L$ and $M_{L}+{M_{J}=M}$ simultaneously. Using the classical interpretation of angular momenta as vectors: in case, when the angular momentum of impurity's internal rotation does not change its absolute value, the induced change of total angular momentum of the impurity  cannot be orthogonal to its external angular momentum. This fact is also in agreement with the geometric constraints that we consider.

\section{Strong-coupling regime}
\label{sec:strong-coupling}

%\oldtext{In the regime of strong coupling between impurity and bath degrees of freedom, we consider the Hamiltonian in the laboratory frame (\ref{eq:hamiltonian}) and apply the Pekar approach~\cite{Pekar}.}

In the strong-coupling regime the deformation of the bath by the presence of the impurity is macroscopic, hence the problem can no longer be approached by considering a single-phonon variational ansatz (\ref{eq:ansatz}). Therefore, in this regime we consider the Hamiltonian  (\ref{eq:hamiltonian}) and apply the Pekar approach~\cite{Pekar}.
Similar approaches have been used for the polaron~\cite{Pekar-polaron} and angulon~\cite{PRA-li} quasiparticles as well as for the rotating polaron~\cite{PRB-polaron}. The trial wavefunction is defined as a tensor product of the impurity state $\ket{\phi_{\text{imp}}}$ and the bath state $\ket{\xi_{\text{bos}}}$:
\begin{equation}
	\ket{\psi}=\ket{\phi_{\text{imp}}}\otimes\ket{\xi_{\text{bos}}} \,,
	\label{eq:pekar-wavefunction}
\end{equation}
and is used to define the Hamiltonian projected on the impurity state
\begin{multline}
	\hat{\mathcal{H}}_{R}=\expval{\hat{\mathcal{H}}}=B_{\text{ext}}\expval{\hat{J}_{z}+\hat{L}_{z}}+B_{\text{0}}\expval{\hat{J}_{z}^{2}}+B\expval{\hat{\bm{L}}^{2}}\\
	+\sum\limits_{k\lambda\mu}\Big(\omega(k)\hat{b}^{\dagger}_{k\lambda\mu}\hat{b}_{k\lambda\mu}+\Big[\expval{\hat{V}_{k\lambda\mu}(\hat{\varphi}_{r},\hat{\Omega})}\hat{b}^{\dagger}_{k\lambda\mu}+H.c.\Big]\Big) \,,
	\label{eq:reduced-hamiltonian}
\end{multline}
where $\expval{\hat{A}}\equiv\bra{\phi_{\text{imp}}}\hat{A}\ket{\phi_{\text{imp}}}$ and the interaction matrix element
\begin{equation}
	\expval{\hat{V}_{k\lambda\mu}(\hat{\varphi}_{r},\hat{\Omega})}=\expval{\sum\limits_{l\delta\alpha\gamma}\mathcal{U}^{\delta\gamma\mu}_{l\alpha\lambda}(k)\hat{Y}^{\ast}_{l,\delta}\Big(\frac{\pi}{2},\hat{\varphi}_{r}\Big)\hat{Y}^{\ast}_{\alpha,\gamma}\big(\hat{\Omega}\big)} \,.
\end{equation}

The projected Hamiltonian (\ref{eq:reduced-hamiltonian}) can be diagonalised with respect to bosonic operators via the following unitary coherent-state transformation
\begin{equation}
	\hat{U}=\exp\Bigg[-\sum\limits_{k\lambda\mu}\frac{1}{\omega(k)}\bigg(\expval{\hat{V}_{k\lambda\mu}(\hat{\varphi}_{r},\hat{\Omega})}\hat{b}^{\dagger}_{k\lambda\mu}-H.c.\bigg)\Bigg] \,.
\end{equation}
The energy functional is subsequently obtained by taking the expectation value with respect to the ground state of bosonic bath, $\ket{0}$,
\begin{multline}
	\mathcal{E}_{R}=\bra{0}\hat{U}^{-1}\hat{\mathcal{H}}_{R}\hat{U}\ket{0}=B_{\text{ext}}\expval{\hat{J}_{z}+\hat{L}_{z}}+B_{\text{0}}\expval{\hat{J}_{z}^{2}}\\
	+B\expval{\hat{\bm{L}}^{2}}-\sum\limits_{k\lambda\mu}\omega^{-1}(k)\Big|\expval{\hat{V}_{k\lambda\mu}(\hat{\varphi}_{r},\hat{\Omega})}\Big|^{2} \,.
\end{multline}
In other words, we have chosen the bosonic state to be $\ket{\xi_{\text{bos}}}=\hat{U}\ket{0}$. Such state accounts for an infinite amount of phonon excitations summarized in a coherent way, i.e. it is the coherent state of the bath.

In general, the impurity state is given by a superposition of the angular momentum eigenstates
\begin{equation}
	\ket{\phi_{\text{imp}}}=\sum\limits_{jmn}\beta_{jmn}\ket{n}\ket{jm} \,,
\end{equation}
which mixes external $\ket{n}$ and internal $\ket{jm}$ rotational states. The energy functional may therefore be written as a function of the variational coefficients $\beta_{jmn}$
\begin{multline}
	\mathcal{E}_{R}=\sum\limits_{jmn}|\beta_{jmn}|^{2}[B_{\text{ext}}(n+m)+B_{\text{0}}n^{2}+Bj(j+1)]\\
	-\sum\limits_{k\lambda\mu}\frac{1}{\omega(k)}\Bigg|\sum\limits_{\substack{jmn \\ j^{\prime}m^{\prime}n^{\prime}}}\beta^{\ast}_{j^{\prime}m^{\prime}n^{\prime}}\beta_{jmn}\tilde{\mathcal{U}}^{k\lambda\mu}_{jmn,j^{\prime}m^{\prime}n^{\prime}}\Bigg|^{2} \,,
	\label{eq:pekar-energy}
\end{multline}
with the coupling
\begin{multline}
	\tilde{\mathcal{U}}^{k\lambda\mu}_{jmn,j^{\prime}m^{\prime}n^{\prime}}=\sum\limits_{\alpha l}U_{\alpha}(k)\sqrt{\frac{(2\alpha+1)^{2}(2l+1)(2j^{\prime}+1)}{(2j+1)(2\lambda+1)}}\\
	\cdot i^{\lambda-\alpha-l}j_{l}(kr_{0})Y_{l,\delta}\Big(\frac{\pi}{2},0\Big)C^{j0}_{\alpha 0,j^{\prime}0}C^{jm}_{\alpha\gamma,j^{\prime}m^{\prime}}C^{\lambda 0}_{\alpha 0,l0}C^{\lambda\mu}_{\alpha\gamma,l\delta} \,,
\end{multline}
where $\gamma=m-m^{\prime}$ and $\delta=\mu-m+m^{\prime}$ subjected to the restrictions of the Clebsch-Gordan coefficients.

Provided that the emergent interaction between external and internal angular momenta in the weak-coupling regime is a consequence of a phonon exchange, we would expect to observe a similar, yet more pronounced, effect, considering the ansatz with the multitude of phonons. The correlation can be written in terms of the variational coefficients as

\begin{equation}
	\mathcal{C}=\sum\limits_{jmn}mn\big|\beta_{jmn}\big|^{2}-\Big(\sum\limits_{jmn}m\big|\beta_{jmn}\big|^{2}\Big)\Big(\sum\limits_{jmn}n\big|\beta_{jmn}\big|^{2}\Big) \,.
	\label{eq:pekar-correlation}
\end{equation}

\begin{figure}
    \centering
    \vspace{0.5\baselineskip}
    \includegraphics[width=8.6cm]{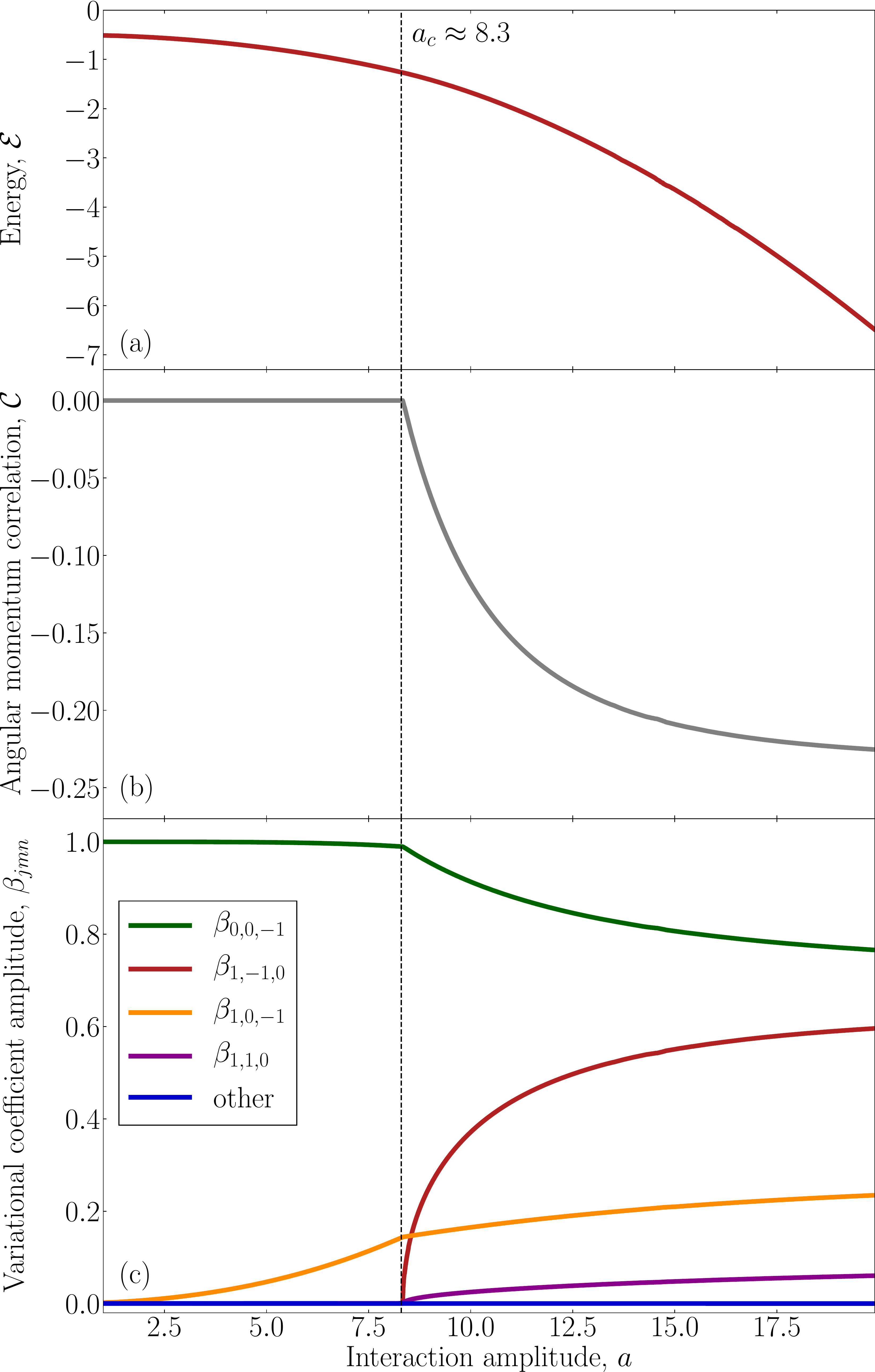}
	\vspace{-0.5\baselineskip}
    \caption{(a) Ground state energy dependence on the impurity-bath interaction strength. The values are obtained through the minimization of Pekar functional \eqref{eq:pekar-energy}. (b) Angular momentum correlation dependence on the  interaction strength. At the critical value $a_{c}\approx 8.3$, the system exhibits transition to the spin-orbit coupled regime. (c) Dependence of the variational coefficients amplitudes on the interaction amplitude. The transition to the correlated regime is explained by the redistribution of coefficients' amplitudes at values higher than the critical strength $a_{c}$.}
    \label{fig:pekar}
\end{figure}

For the numerical evaluation of Eqs.~\eqref{eq:pekar-energy} and \eqref{eq:pekar-correlation}, we use a simplified model of interaction potential between the impurity and bath degrees of freedom:
\begin{equation}
	U_{\lambda}(k)=\frac{ak}{\sqrt{2\lambda+1}}e^{-k} \,,
\end{equation}
with $a$ being a free parameter defining the strength of interaction. This model potential resembles the behavior of Eq.~\eqref{eq:interaction-potential} and together with the simplified dispersion relation $\omega(k)\equiv 1$, it effectively decreases the complexity of computations. 

The Pekar energy functional \eqref{eq:pekar-energy} is minimized with respect to the variational coefficients to define the energy of the ground state, which is presented in Fig.~\ref{fig:pekar}(a). Energy values exhibit a smooth behavior and experience a decay with increasing interaction strength. In Fig.~\ref{fig:pekar}(b) we show the angular momentum correlation dependence on the interaction amplitude. The correlation function, similarly to Ref.~\cite{PRA-li}, exhibits a sharp transition to non-zero values at the critical point $a_{c}\approx 8.3$ and shows a saturation at high values of $a$. To explain this behavior, in Fig.~\ref{fig:pekar}(c) we show the dependence of variational coefficients $\beta_{jmn}$ amplitude on the interaction strength. In the low impurity-bath interaction regime, the system is in its ground state $\ket{0,0,-1}$ with the first term in Eq.~\eqref{eq:pekar-energy} dominating. At the transition threshold $a_{c}$, however, the exchange between internal and external rotations occurs and the system exhibits the states that maximize the interaction with the surroundings: $\ket{1,-1,0}$,$\ket{1,0,-1}$,$\ket{1,1,0}$. Consequently the second (interaction) term in Eq.~\eqref{eq:pekar-energy} starts prevailing, manifesting the correlation. It is worth noticing that the energy of the system, depicted in Fig.~\ref{fig:pekar}(a), does not show a kink, implying that the transition between different angular momentum configurations occurs within the same energy level.

These results explicitly connect the angular momentum exchange between internal and external rotations with the correlation of the angular momentum operators. Since the latter was chosen as a measure of the spin-orbit coupling in the system, the strong-coupling approach indicates that such interaction is mediated by excitations in many-body environment. 

It is worth pointing out that the strong- and weak-coupling regimes consider the opposite limits of the coupling strength between the impurity degrees of freedom and the bath. The deformations of the bath induced by the presence of a molecular impurity in different regimes are thus on different scales. Consequently, none of the trial wavefunctions \eqref{eq:ansatz} and \eqref{eq:pekar-wavefunction} can be represented as a limiting case of the other. Although the results of Sections \ref{sec:weak-coupling} and \ref{sec:strong-coupling} are in agreement with each other, they cannot be deduced from one another and each of the approaches is required to examine the problem in a more general way.

\section{Conclusions} \label{sec_conc}

To summarize, we have obtained the effective Hamiltonian of a rotating mobile impurity immersed in a bosonic medium, in the case where the center-of-mass motion is confined to a ring. We calculated the spectral function of the system in the regime of weak coupling between the molecule and the many-body environment. We observed transitions between distinct energy levels occuring in the vicinity of the angulon instability \citep{PRL-angulon}. The transition probabilities were found to be dependent on the relative angular momenta orientations in system. This leads to a conclusion that the energy of each state depends on the coupling between external and internal rotations, similar to the conventional spin-orbit interaction. We have used the correlation between angular momentum operators as a quantitative measure of such coupling. The numerical calculation has shown that the amplitude of this correlation depends strongly on the relative orientations of the angular momenta. This allowed us to connect the excitations in the phonon continuum observed in the spectrum with the interaction between external and internal rotational degrees of impurity.

To provide a broader view on the problem, we applied a Pekar approach to the regime, where the impurity is strongly interacting with the medium. In this case we considered a coherent state of phonon excitations, instead of the single-phonon variational ansatz used in the weak-coupling regime. Thus, the spin-orbit interaction was expected to be more pronounced. The numerical study revealed that the main factor inducing the coupling between internal and external degrees of freedom is the redistribution of the state amplitudes above the critical interaction strength. Since the moment when the interaction with surroundings starts dominating, the system chooses to occupy the states that maximize this interaction. This induces the correlation between external and internal angular momenta. This result implies the existence of an effective spin-orbit interaction mediated by the many-body environment. The sharp transition to the interacting regime is promising for the applications either directly in few-impurity systems in the presence of a medium (similar to Ref.~\cite{SO-laser-1,SO-laser-2,SO-laser-3}) or in a large set of collective phenomena that arise as a consequence of such interactions, i.e., spintronics phenomena~\cite{Rashba-spintronics,Spintronics-review} or topological electronic states~\cite{TI-1,TI-2,TI-3}.

\section*{Acknowledgements}
We are grateful to G. Bighin and A. Ghazaryan for valuable discussions. M.L and E.Y.~acknowledge support by the Austrian Science Fund (FWF), under project No.~P29902-N27, and by the European Research Council (ERC) Starting Grant No.~801770 (ANGULON).

\bibliographystyle{apsrev4-1} % Tell bibtex which bibliography style to use
\bibliography{bibliography.bib}

\end{document}